# A New Approach to Detecting and Designing Living Structure of Urban Environments

Bin Jiang[1] and Ju-Tzu Huang[2]

[1]Faculty of Engineering and Sustainable Development, Division of GIScience
University of Gävle, SE-801 76 Gävle, Sweden
Email: bin.jiang@hig.se

[2]Department of Real Estate and Built Environment
National Taipei University, 237 New Taipei, Taiwan
Email: rutszhuang@gmail.com

*In such a world, scientists would do better, the profound questions of health, wholeness, nature, ecology, and human joy, would be part of a single world view, in which it would be recognized as part of science – scientia – that is to say, knowledge – and in which scientists and artists together, speaking a common language, would take part in this joy, to the benefit of all humankind.*

Christopher Alexander (2003)

**Abstract**
Sustainable urban design or planning is not a LEGO-like assembly of prefabricated elements, but an embryo-like growth with persistent differentiation and adaptation towards a coherent whole. The coherent whole has a striking character – called living structure – that consists of far more small substructures than large ones. To detect the living structure, natural streets or axial lines have been previously adopted to be topologically represent an urban environment as a coherent whole. This paper develops a new approach to detecting the underlying living structure of urban environments. The approach takes an urban environment as a whole and recursively decomposes it into meaningful subwholes at different levels of hierarchy or scale ranging from the largest to the smallest. We compared the new approach to natural street and axial line approaches and demonstrated, through four case studies, that the new approach is better and more powerful. Based on the study, we further discuss how the new approach can be used not only for understanding, but also for effectively designing or planning the living structure of an urban environment to be more living or more livable.

## 1. Introduction
Sustainable urban design or planning is not an assembly of prefabricated elements, like LEGO, but an embryo-like growth with persistent differentiation and adaptation towards a coherent whole (Alexander 1987, 2002–2005). The coherent whole is also called living structure or wholeness, which is a physical and mathematical structure that consists of many substructures with an inherent hierarchy (Alexander 2002–2005, Jiang 2019). Across different levels of the hierarchy, there are far more small substructures than large ones (or scaling law) (Jiang 2015a, Salingaros and West 1999), whereas on each level of the hierarchy, there are more or less similar substructures, so called Tobler's law. It is commonly referred to as the first law of geography, which charmingly states that *everything is related to everything else, but near things are more related than distant things* (Tobler 1970). Living structure exists pervasively in our surrounding. For example, a country is a living structure because it consists of far more small cities – or substructures in general – than large ones across all scales, while cities on each scale tend to be more or less similar. A city is a living structure because it consists of far more short streets – or substructures in general – than long ones across all scales, while streets on each scale tend to be more or less similar. Thus, both country and city are living structures, and can therefore be said to have a high degree of life or beauty. To detect living structure, we can represent a city into numerous streets identified by individual names and good continuity or axial lines representing the longest visibility lines for individual spaces (Hillier and Hanson 1984, Jiang et al. 2008). Both streets and axial lines of an urban environment constitute a coherent whole or living structure. However, these approaches are not particularly instructive or straightforward for helping sustainable urban design or planning.

The present paper develops a new approach for efficiently detecting the underlying living structure and its substructures, and thereafter for effectively designing livable urban environments. The approach takes an urban environment as a whole and decomposes it into numerous subwholes in a step-by-step fashion. The whole can be said to be persistently – in a recursive manner – differentiated to numerous subwholes (so called differentiation principle) with far more smalls than larges across scales, whereas the resulting subwholes are well adapted to each other (so-called adaptation principle) with more or less similar on each scale. This approach is developed under the third view of space, or an organismic cosmology, first conceived by Whitehead (1929) and further developed by Alexander (2002–2005) that space is neither lifeless nor neutral, but a living structure capable of being more living or less living. Under this organismic world view, the world is a coherent whole rather than fragmented pieces, as currently conceived and perceived under the Cartesian world picture (Descartes 1637, 1954). For example, a house is a living structure, which is a substructure of a larger living structure of the street, which is a substructure of a larger living structure of the city, which is a substructure of a larger living structure of the country, and so on towards the entire Earth's surface. Theoretically speaking, the Earth's surface is the only living structure that geography is concerned about, and all others smaller than the Earth's surface are called its substructures. This holistic view of space implies that any design or planning action at any level of scale would affect virtually the livingness of the entire Earth's surface.

This paper is further motivated by the state of the art of science, or urban science in particular, which is largely concerned about understanding complexity rather than creating various complex or living structures. This state of science applies not only to conventional science conceived as separated from art and humanities (Snow 1959), but also to complexity science (Simon 1962, Alexander 2003) such as fractal geometry (Mandelbrot 1983), complex networks (Newman 2006), and generative science (Epstein 1999, Wolfram 2003), all related to complex or beautiful patterns or structures. However, none of these sciences really intend to address the issue of creating structures. In some cases, they do involve creating structures, but they are unable to address why a structure is beautiful, and how much beauty a structure is. Instead, the issue of creating structures or things is commonly left to art and ascribed to a matter of opinions or personal preferences. For example, fractal geometry is able to generate fractal patterns, but when it comes to how good or how beautiful a fractal pattern is, it has no formal answer. It is the same for cellular automata patterns that emerged from a set of transitional rules (Wolfram 2003). The notion of living structure directly confronts the issue of beauty and aims not only to better understand complexity, but also to effectively create beautiful or living structures (Alexander 2003, Salingaros 2012, 2013, Jiang 2015b, Mehaffy 2017, Gabriel and Quillien 2019). More importantly, the created living structure can be judged objectively in terms of its degree of beauty or livingness; the higher the degree of living structure, the more beautiful a structure is (Jiang 2019). Therefore, living structure is to beauty what temperature is to warmness.

This paper makes three major contributions to the literature. First, living structure is stated as the structure of the structure of the structure and so on, or a living structure consists of numerous substructures with far more smalls than larges. Second, the new approach is developed for detecting the living structure of urban environments, not only for understanding but also for effectively designing livable urban environments. Third, four case studies are conducted by applying the new approach to urban areas from France and UK to verify the developed approach and why it is better than the axial line and natural street approaches.

The remainder of this paper is organized as follows. Section 2 introduces the concept of living structure from a dynamic or design point of view and shows how a living structure is transformed or differentiated in a step-by-step fashion. Section 3 presents the new approach to detecting living structure and its substructures at different levels of hierarchy or scale by a working example of leaf vein. In Section 4, the new approach is further verified and compared with natural street and axial line models, to demonstrate that the new approach is better and more powerful for understanding and designing living structure of urban environments. Section 5 further discusses the new approach and its implications on sustainable urban design. Finally, Section 6 concludes this paper and points to future work.

**2. A city as a transformed living structure and the degree of livingness or beauty**
The most profound design thought advocated by Alexander (1987, 2002–2005) and his associates such as Salingaros (2005) and Mehaffy (2017) is that a city is a coherent whole that emerged and evolved in a piecemeal fashion rather than a LEGO-like assembly of prefabricated elements. In other words, a city grows out of wholeness, another name of living structure, in a step-by-step style, very much like an embryo that grows to be a human being. This is the reason why most beautiful towns and cities of the past – such as Venice and Amsterdam – are able to convey a feeling of unity or wholeness or being organic. This feeling of wholeness appears at different levels of hierarchy or scale, not only at the largest scale of city, but also in a series of small scales: in neighborhoods, in streets, in restaurants, in sidewalks, in houses, shops, markets, parks, gardens, walls, even in balconies and ornaments. Importantly, this feeling of wholeness or being organic is neither a vague feeling of relationship with biological forms nor an analogy. Instead, wholeness or living structure is a physical and mathematical structure that is governed by the two fundamental laws as mentioned at the beginning of this paper. Let us look at an example of city evolution that Alexander often cited in his books and papers.

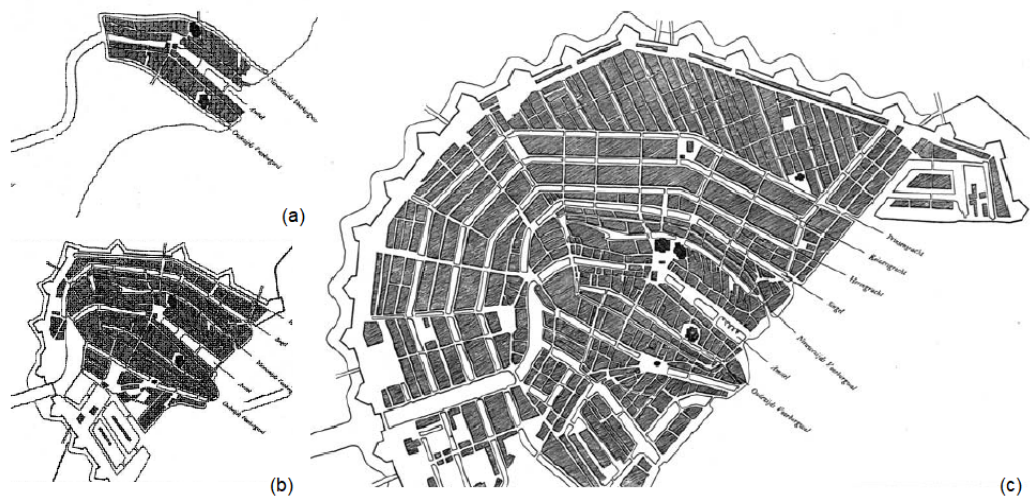

Figure 1: Amsterdam over 500 years up to the late 18th century
(Note: Amsterdam when it was founded in the 14th century (a), organismically evolved in the 15th century (b) and in the late 18th century (c).)

The traditional city of Amsterdam was organismically emerged and evolved for a few hundred years, so it had a very organic shape. Figure 1 shows three stages of Amsterdam in the course of evolution. At the very beginning, back in the 14th century, there was only a U-shaped wall around a few city blocks, with the river Amstel down the middle (Figure 1a). This is the original whole at the first stage. The whole was well adapted to the position and shape of the mouth of the river Amstel. The original whole had a horseshoe-like structure, which was only dimly present around the white part of the drawing in Figure 1a. This horseshoe-like structure became real and clear in the second state of the city (Figure 1b). This larger horseshoe shape had been realized by the surrounding concentric canals and streets, and canals had been built to drain the land, all following the natural line of the originally latent horseshoe. The concentric structure of canals and streets had been intensified in the third stage around the late 18th century, by adding further layers and filling in a much larger area in a way that supports and continues the structure of the second stage.

The example of Amsterdam city may sound a bit abstract, particularly for readers who are unfamiliar with the notion of living structure, or for readers who view a city as a LEGO-like assembly. The key message from this example is that, at every stage, we shall ask whether the current development has grown out of the wholeness that was there before. The wholeness must be preserved, enhanced, and deepened; otherwise, it would not be considered to be a good development. Therefore, this kind of transformation is called wholeness-enhanced transformation (Alexander 2005). Based on the living structure view of city evolution, it is not difficult to understand why a city is essentially unpredictable (Batty 2018), because of the piecemeal or iterative nature of development.

Given two living structures, it is possible to objectively judge which one holds a higher degree of livingness. To make this idea clear and to further illustrate scaling law and Tobler's law, let us examine two paintings (Figure 2) in terms of their degrees of livingness or beauty. The first painting is called *Composition*, by the Dutch painter Piet Mondrian (1872–1944), one of the greatest artists of the 20$^{th}$ century, and famous for his paintings using simple geometry and primary colors. The second painting is modified from *Composition* and it is named *Configuration* for the sake of convenience. These two paintings look very similar to each other, with a slight difference in the left strip. *Configuration* is objectively more living or more beautiful than *Composition,* as illustrated in Figure 2. Under the assembly view, both paintings are composed of seven and nine pieces or substructures. However, under the living structure view, there are in total 18 (7 + 6 + 4 + 1) substructures for *Composition*, whereas there are 20 (9 + 6 + 4 + 1) substructures for *Configuration*. It is important to note that the large scales are embedded in the small scales for the living structure view. Seen from the living structure point of view, *Composition* is less living or less beautiful than *Configuration*, because the notion of far more smalls than large occurs only twice for the former, compared to three times for the latter.

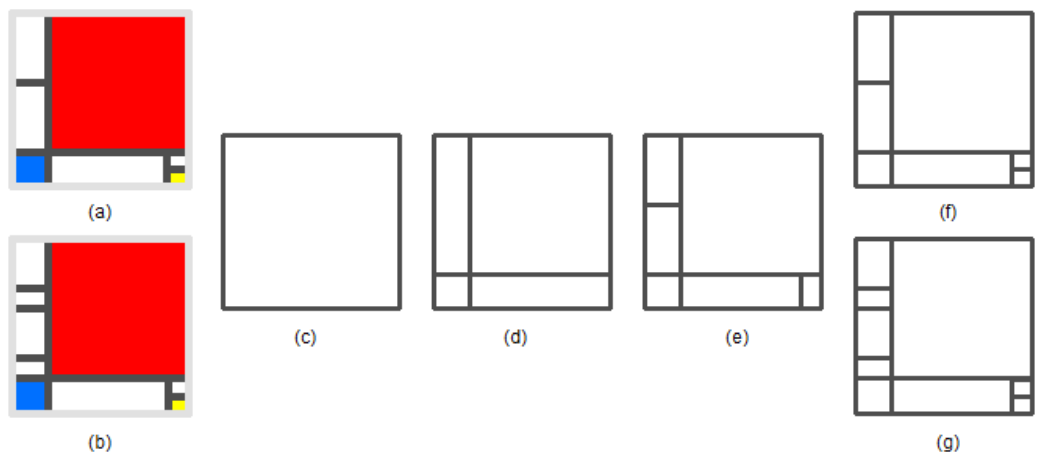

Figure 2: (Color online) Illustration on why one painting is more structurally beautiful than the other (Note: The two paintings – *Composition* (a) by the famous Dutch painter Piet Mondrian (1872–1944), and *Configuration* (b) modified from *Composition* by the first author of this paper – meet the minimum condition of being a living structure. Both paintings can be viewed to be differentiated like cell division from the square, so they are featured by the recurring notion of far more newborn (newly generated) substructures than old ones. More specifically, there are far more newborns (4) than old one (1) from (c) to (d), and again far more newborns (6) than old ones (4) from (d) to (e). However, these paintings differ only in the final transformation. That is, from (e) to (f) there is a violation of far more newborns (7) than old ones (6), because 7 and 6 are more or less similar, whereas from (e) to (g) the notion of far more newborns (9) than old ones(6) remains. On the other hand, in each iteration there are far more small substructures than large ones. Thus, the painting *Configuration* is more living or more beautiful – structurally – than the painting *Composition*.)

Figure 2 presents not only two paintings with different degrees of beauty, but also a recursive or iterative way of seeing things as a living structure. A living structure is a set of differentiated substructures at different levels of hierarchy or scale with far more smalls than larges. In other words, a living structure consists of far more small substructures than large ones. According to Alexander's definition of living structure, the differentiated substructures are called centers, the only building blocks of a living structure. In this paper, we replace the term *center* with *substructure*, which is simpler for better understanding living structure. That is, a living structure is the structure of the structure of the structure and so on. The recursive way of stating living structure also constitutes the foundation of the new approach to detecting living structure.

### 3. A new approach to detecting living structure and its substructures

The new approach is developed based on black and white city plans to automatically detect all of the smallest substructures, from which large substructures at upper levels of hierarchy or scale can be merged in a step-by-step fashion. For the lowest or smallest substructures, we use the head/tail breaks (Jiang 2013a) to have them detected automatically. Before getting into details of the approach, let us first introduce the head/tail breaks and the degree of beauty.

### 3.1 Head/tail breaks and the degree of beauty

Head/tail breaks is not only a classification scheme for data with a heavy-tailed distribution, but also a visualization tool for big data analytics. It is an iterative function, which partitions a data as a whole around the average into the head for those big values and the tail for those small values, and then derives iteratively the head of the head of the head, and so on. To illustrate how head/tail breaks works, let us use the 10 number $[1, 1/2, 1/3, …, 1/10]$ to show how it can be classified (Figure 3). The average of the 10 number is approximately 0.29, which partitions them into two groups: the head for those greater than the average $[1, 1/2, 1/3]$ and the tail for those less than the average $[1/4, 1/5, …, 1/10]$. For the three numbers in the head, the average is 0.61, which further partitions the head into the head $[1]$ and the tail $[1/2, 1/3]$. In general, the partition process continues recursively and for the head, the head of the head of the head, and so on, as long as the length of the subset in the head is less than a preset threshold. For the head/tail breaks 1.0, the length of every head must be less than 40%, whereas for the head/tail breaks 2.0 the average length of all heads must be less than 40%. In this paper, we adopted head/tail breaks 1.0 which simply works well for our purpose.

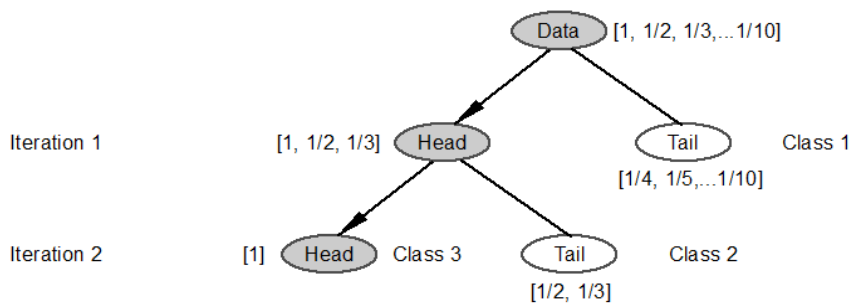

Figure 3: Illustration of head/tail breaks classification with a simple example of the 10 numbers.
(Note: The 10 numbers $[1, 1/2, 1/3, …, 1/10]$ are classified into three classes: $[1/4, 1/5, …, 1/10]$, $[1/2, 1/3]$, and $[1]$, according to the head/tail breaks.)

These 10 numbers also present a very good example for illustrating the notion of living structure. First of all, the 10 numbers follow scaling law, for the notion of far more smalls than larges recurs twice, so the ht-index is 3, leading to the three classes. On the other hand, numbers in each of the three classes are more or less similar, so they follow Tobler's law. The 10 numbers follow Zipf's law (1949) exactly rather than statistically; that is, the first largest city is twice as big as the second largest, three times as big as the third largest, and so on. The essence of Zipf's law is its statistical rather than exact nature. Strictly speaking, the 10 numbers do not follow Zipf's law, whereas its variants $[1 + e_1, 1/2 + e_2, 1/3 + e_3, …, 1/10 + e_{10}]$ (where $e_1, e_2, e_3, … e_{10}$ are very small values) do follow Zipf's law. It is in this sense that the varied 10 numbers are more living or more beautiful – structurally – than the initial 10 numbers.

### 3.2 A working example based on a leaf vein

We use a leaf vein as a working example to introduce the new approach to detecting living structure and its substructures (c.f. Appendix A for a simple tutorial based on ArcGIS). This working example extends the above example of the 10 numbers to one million numbers. Assuming that the leaf vein is with a color image of one million pixels (Figure 4a), we first convert the color image to a gray-scale image (Figure 5a) according to the formula: Gray = 0.299 * Red + 0.587 * Green + 0.114 * Blue, which is commonly used the image processing literature (Poynton 2003). Thus, there are one million pixels with gray scales between 0 and 255. Note that these one million pixels do not follow a heavy-tailed distribution, indicating that the pixel perspective is unable to reveal the living structure. Regardless of this, we still can apply the head/tail breaks to the gray image to get multiple averages recursively (Table 1), from which we chose a meaningful value (171) somewhere around the second mean for delineating the smallest substructures (Figure 4f). From these smallest substructures, large substructures at different levels of scale or hierarchy can be created by merging the small scales (Figure 4b–4e). The outcomes shown in Figure 4 reflect very well what we human beings perceive about the inherent hierarchy of the leaf vein. The spectral coloring is adopted to show the five levels of hierarchy or scale: red for the largest scale, blue for the smallest scale, and the other colors for the scales between the largest and the smallest. The same coloring will be used in other related figures of the paper.

Table 1: Head/tail breaks statistics for the gray image or the one million pixels
(Note: The one million pixels do not follow a heavy-tailed distribution, which indicates that the pixel perspective is unable to reveal the living structure. However, the head/tail breaks still can be used to obtain the meaningful cutoff around the second mean to derive the smallest substructures.)

| #Pixels | #head | %head | #tail | %tail | Mean |
|---|---|---|---|---|---|
| 1000000 | 523895 | 52% | 476105 | 48% | 125 |
| 523895 | 344885 | 66% | 179010 | 34% | 169 |
| 344885 | 248693 | 72% | 96192 | 28% | 181 |

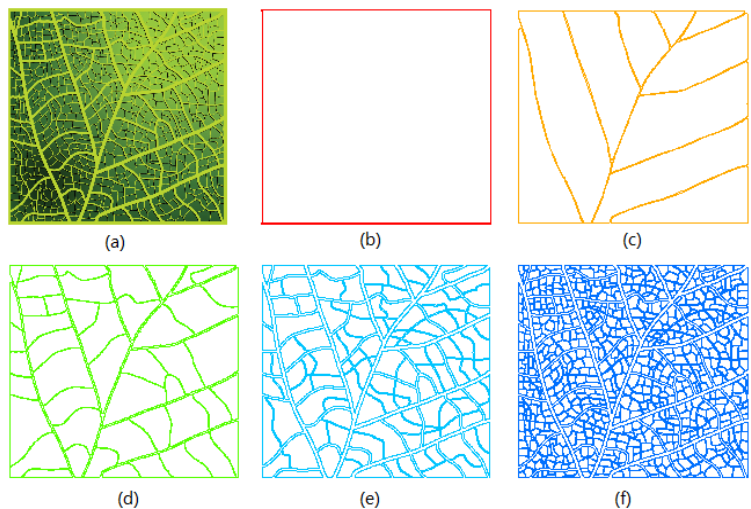

Figure 4: (Color online) A leaf vein (a) is decomposed into many substructures (b–f).
(Note: The leaf structure has five levels of scale (or classes) and is governed by two fundamental laws: scaling law or far more small substructures than larges across all scales from the largest (Panel b in red) to the smallest (Panel f in blue); and Tobler's law or more or less similar sized substructures on each scale (one of these Panels b–f). For example, there are far more smalls (10) than large one (1) from Panel b to Panel c, and there are far more smalls (48) than larges (10) from Panel c to Panel d. The leaf structure can also be considered to be differentiated from a square space with adaption on each scale, thus the two principles of differentiation and adaptation. A fundamental difference between (a) and (f) is that the former has all five levels of scale together, while the latter only has the smallest scale.)

The leaf vein and its inherent hierarchy effectively reflect the two laws of living structure. First, there is the recurring notion of far more smalls than larges across all five scales (Panels b–f of Figure 4), i.e., recurring four times so the ht-index is five, leading the five classes. Second, on each of the given scales, the substructures are more or less similar. From a design point of view, the leaf vein is differentiated from the square into numerous substructures, which are well adapted to each other. Seen from these two laws, the leaf vein is living or beautiful. This paper is about sustainable urban design, so we must be very clear about the structural nature of beauty and the two laws of beauty.

The beauty arises out of the living structure with far more small substructures than large ones. Let us compare the leaf vein with the Sierpinski carpet (1915) as shown in Figure 5. Both have almost the same number of hierarchical levels, but the Sierpinski carpet is less living or less beautiful than the leaf vein. There are two reasons why the Sierpinski carpet is less living: (1) the ratio of square sizes across scales (1/3, 1/9, 1/27, and 1/81) is exactly one third, so too restricted in terms of the scaling law, and (2) the square size on each scale is exactly the same rather than more or less similar in terms of Tobler's law. It is the same issue as we remarked above on Zipf's law; the 10 numbers [1, 1/2, 1/3, …, 1/10] are too restricted to follow Zipf's law. Like Zipf's law, both scaling law and Tobler's law should be correctly understood as statistical regularity. In this connection, strictly speaking, the Sierpinski carpet violates Tobler's law, for the substructures at each scale of the carpet are precisely the same rather than more or less similar. The Sierpinski carpet also violates scaling law, as the notion of far more smalls than larges is too restricted.

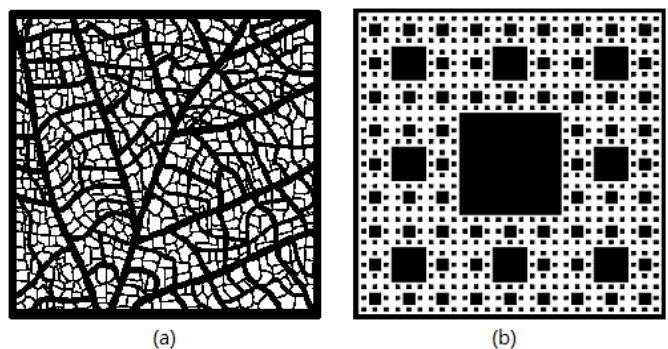

Figure 5: The leaf vein is structurally more living or more beautiful than the Sierpinski carpet
(Note: The leaf vein (a) and the Sierpinski carpet (b) meet scaling law with the same recurring notion of far more smalls than larges, so both are living or beautiful structures. However, the leaf vein has a higher degree of livingness or beauty than the Sierpinski carpet, because the squares on each scale are precisely the same rather than more or less similar, thus violating Tobler's law to some extent. Another obvious reason that the Sierpinski carpet is less beautiful is that its negative space is not well shaped although its positive space is well shaped, by which we meant shapes are convex rather than concave. In contrast, both positive and negative spaces of the leaf vein are well shaped.)

**4. Case studies**
We applied the new approach to four urban environments to further verify it and examine why it is better than the axial line and natural street approaches. The four urban environments include the village of Gassin and the town of Apt, both in south France, and two neighborhoods or districts in London: The city of London and Stockwell Station. These four cases have been well studied in the literature of space syntax (e.g., Hillier 1989, Hillier et al. 1997). It was demonstrated that the beady-ring structure of numerous convex spaces of the Gassin village represents the kind of living structure (Alexander 2002–2005). Figure 6 shows their overall plans, which all look very organic. Note that we have deliberately not shown map scales in these plans, as we concentrated on their overall configurations rather than geometric details such as sizes and directions.

We applied the new approach to detect substructures for the four urban environments. Table 2 reports the number of substructures at different levels of scale, indicating the recurring notion of far more small substructures than large ones. The hierarchal levels vary from four to six, while the number of smallest structures changes from 26 to 462. It is not surprising that the two neighborhoods in London are more living or more beautiful than the two in France. Note that we are talking in structural terms, based on the underlying plans only, without considering the vertical dimension of these urban environments.

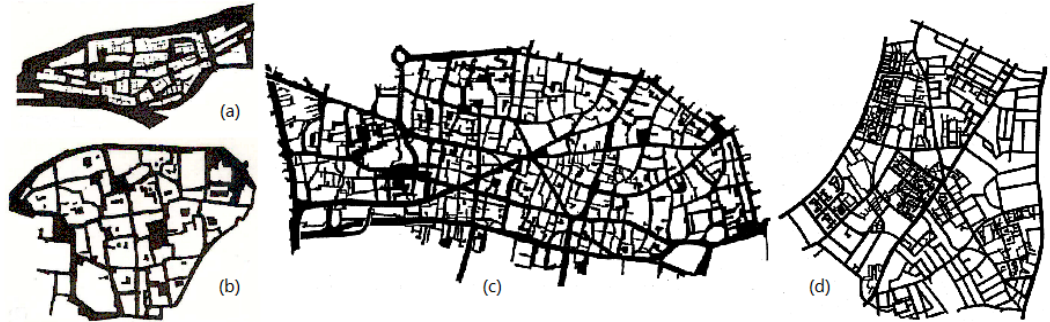

Figure 6: Four case study areas and their plans
(Note: The village of Gassin (a) and the town of Apt (b), both in south France, the city of London (c), and Stockwell Station (d), in central and south London, respectively.)

The overall shape of the city of London looks like a fish, but the organic or living nature of the environment lie in its substructures, with far more smalls than larges. For these substructures, the notion of far more smalls than larges occurs or recurs five times, so the ht-index is six, leading to the six classes. In other words, the City of London plan is a living structure consisting of far more small substructures than large ones. Figure 7 shows the six levels of scale of substructures, ranging from the largest (red) to the smallest (blue), represented by the spectral coloring suggested earlier. The inherent hierarchy of living structure can be expressed in another way: the numerous smallest (blue), the one largest (red), and some in between the smallest and the largest (the other colors between red and blue). The recurring notion of far more smalls than larges can also be reflected in the power law distribution between the size of and the number of substructures, as shown in Figure 8.

Table 2: The number of substructures (SS) at different levels of scale

| Place | #SS | #SS | #SS | #SS | #SS | #SS | Ht-index |
|---|---|---|---|---|---|---|---|
| Gassin Village | 26 | 15 | 7 | 3 | 1 | | 5 |
| Apt Town | 42 | 9 | 3 | 1 | | | 4 |
| City of London | 383 | 33 | 10 | 5 | 2 | 1 | 6 |
| Stockwell Station | 462 | 28 | 9 | 4 | 2 | 1 | 6 |

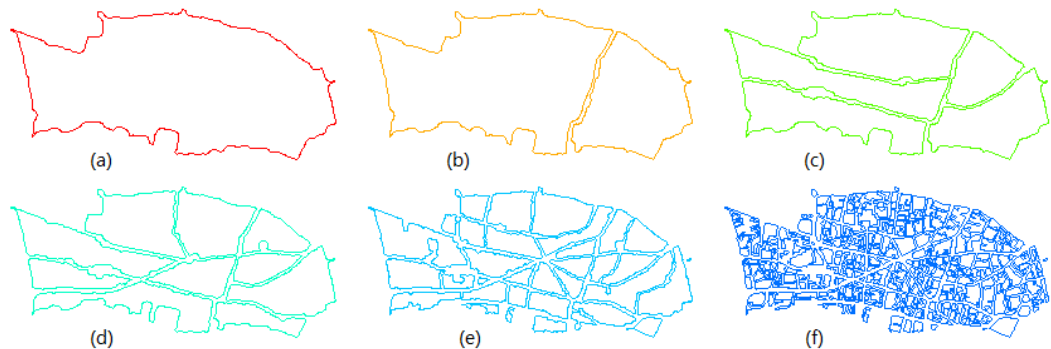

Figure 7: (Color online) The substructures at six levels of scale for the city of London
(Note: The city of London plan is obviously a living structure, as it meets scaling law or the recurring notion of far more smalls than larges across a range of scales from the largest to the smallest (a–f), and Tobler's law or more or less similar substructures on each of the scales indicated by each color. All smaller scales are more living or more beautiful than the larger scales. On the other hand, the smallest scale can be further differentiated in the future to make it even more living or more beautiful.)

The detected substructures at the different levels of hierarchy or scale make a better sense from a design point of view. The substructures are differentiated from scratch, an empty space in Figure 7a. The empty space is differentiated or transformed by persistently increasing its degree of unity or wholeness in a step-by-step fashion. In other words, the empty space is filled with detailed or small structures, which makes it more and more living, or more and more beautiful. Every level of subsequent structures (or panel) is more living than the previous level (or panel) in Figure 7, and the structure in Panel (f) is the most beautiful among the others. The city of London seems well developed towards a very beautiful structure, while part of the Stockwell Station can be developed further. This is the piecemeal design thought that has been advocated by Alexander (1987, 2002–2005) and Salingaros (2005), among many others. Now, with the new approach, we can not only diagnose an urban environment in terms of its degree of livingness, but also effectively re-design it towards more living or more beautiful.

The new approach takes a holistic perspective on an urban environment and is therefore more effective than natural street and axial line models. Both natural street and axial line models are still fragmented, despite their connected graphs that are seemingly linked as wholes. Let us qualitatively compare the new approach to the natural street and axial line models (Jiang et al. 2008, Hillier and Hanson 1984). First, we create a simulated natural street model out of the substructures based on the city of London. With reference to Figure 7, let us assume a ring road (in red) surrounding the city of London, which makes it the longest or the most important street. The second important street would be the one induced by the two largest substructures, the linear space between these two substructure in Figure 7b, followed by the third important streets, the three streets induced by the five substructures in Figure 7c, and so on. These induced natural streets would have the same inherent hierarchy as that of the substructures (Figure 9). To this point, the reader would have little difficulty understanding the similarity between the inherent hierarchy of the substructures and that of the induced natural streets.

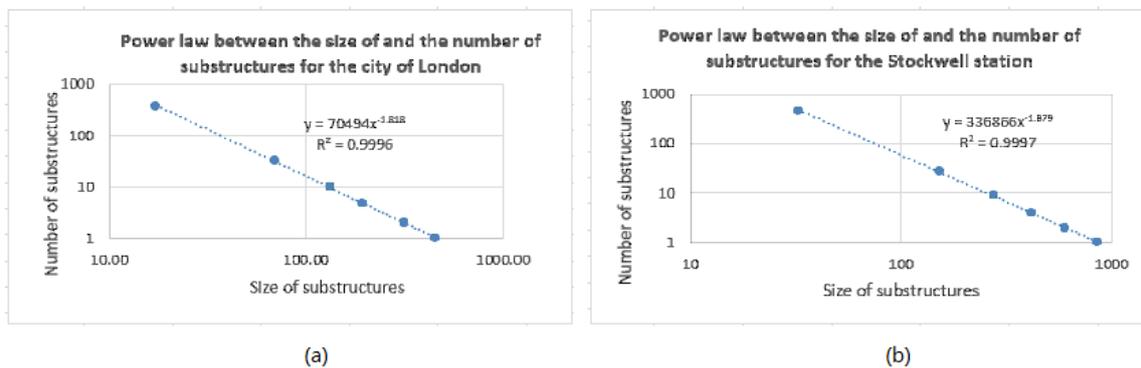

Figure 8: (Color online) Power laws between the size and number of substructures

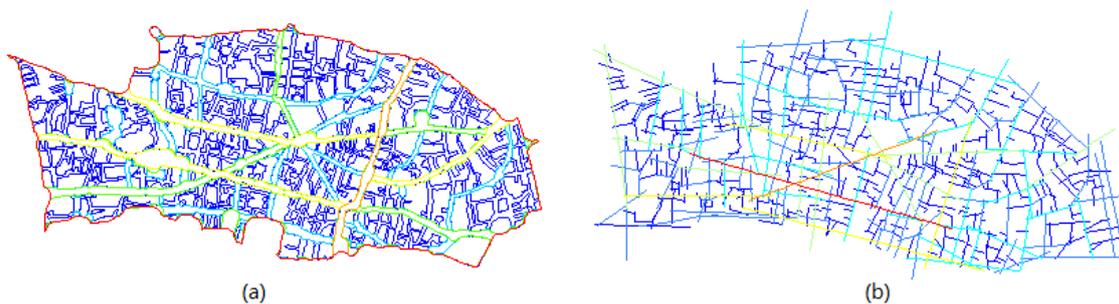

Figure 9: (Color online) Simulated natural streets (a) and manually drawn axial lines (b)
(Note: The simulated natural streets are induced by the substructures, so they have the same degree of hierarchy as that of the living structure of the substructures, as shown in Figure 7. Assuming a ring road bounded the urban environment, it has the highest degree of connectivity show in red. The most connected ring road (red) is followed by the brown street, induced by two brown substructures, and then three yellow-greenish streets, and so on. Truly natural streets would look more like axial lines (b), although the natural streets would be less fragmented than the axial lines.)

If the natural street and axial line models were the same as the simulated natural street model, they would have the same effect as the living structure. In fact, both natural street and axial line models are more fragmented than the simulated one (Figure 9). For example, the ring road is just hypothesized, and it consists of many natural streets (Jiang et al. 2008). These natural streets with other natural streets would have to be chopped into even more axial lines depending on the curvature of the natural streets. Apparently, the axial lines would be more fragmented than the natural streets. We can clearly see from the comparison that the new approach is better than the natural street and axial line models, both in terms of understanding and designing the living structure of an urban environment. The simulated natural streets are the gaps between different substructures of different levels of hierarchy or scale, so they are parts of the substructures or living structure in general under the organismic way of thinking. On the other hand, actual natural streets are usually conceived as paths under Cartesian mechanistic thinking.

It is important to stress the recursive nature of living structure. For example, the smallest substructures (as a whole) shown in Figure 7f, or those shown in Figure 4f, looks exactly as the original pattern (Figure 6c or Figure 4a). However, we must realize that both Figure 6c and Figure 4a contains different levels of hierarchy or scale. It is this living structure or recursive perspective that makes Figure 6c different from Figure 7f, and Figure 4a different from Figure 4f. It is also this living structure or recursive perspective that makes the growth design thought different from the LEGO-like assembly thought.

When applying the new approach for sustainable urban design or planning, substantial consideration should be given to a larger area instead of its own area alone, which is part of the larger area. Based on the notion of wholeness or living structure, goodness of a space relies not only on its own substructures or subspaces, but also on a larger space that contains the space. The larger space should be governed by two fundamental laws, implying (1) far more small substructures than large ones across scales, and (2) more or less similar substructures on each scale. A simple solution would be to apply the approach to a substantially larger area, of which the to-be-designed area is a substructure of the larger living structure.

### 5. Implications of the new approach on sustainable urban design or planning
The new approach to detecting living structure begins with a space and decomposes it, according to its inherent hierarchy, into many substructures at different levels of hierarchy or scale with far more smalls than larges. These decomposed substructures enable us to see not only why an urban environment is living, but also how much living it is: the more hierarchal levels, the more living or more beautiful. We have already seen how the new approach can help not only achieve a better understanding of living structure of an urban environment, but also make the urban environment more living or more beautiful. In this section, we will further discuss the new approach and its implications on sustainable urban design.

While many urban theories have been developed in the past for understanding how cities structure and function, none has really aimed for urban design or planning. Even worse, urban theories have been accused of being pseudoscientific (e.g., Cuthbert 2007, Marshall 2012) or based on *"a foundation of nonsense"* (Jacobs 1961). The new approach or the theory of living structure in general provides a larger theory that can help justify previous urban theories. For example, underlying the axial lines and natural streets is essentially a living structure that makes human movement predictable; this point has also been made clear in the above case studies. It is the underlying living structure that constitutes the image of the city (Lynch 1960, Jiang 2013b). The detected substructures at different levels of hierarchy or scale constitute organized complexity, as advocated by Jacobs (1961), and fundamentally resemble the model of central place theory (Christaller 1933, 1961) in a statistical rather than strict manner (Jiang 2018). It is the theory of living structure that was developed initially for effectively designing or planning cities. Both geography and urban science should go beyond the understanding of complexity, towards making or remaking living or more living structures. This is the new kind of science that Alexander (2003) has advocated, where both scientists and artists share a common language about the goodness of things.

There are three fundamental issues about a city or a geographic space in general: (1) how it looks, (2) how it works, and (3) what it ought to be. The two fundamental laws – scaling law and Tobler's law – can help address the first two issues, whereas the two design principles – differentiation and adaptation – can help make a space or a city to be more living or more beautiful. What underlies the two laws and the two principles is the organismic way of thinking that takes our surroundings (such as gardens, buildings, streets, cities, landscapes, and countries) as substructures of the Earth's surface. Any design or planning action would contribute to the beauty of the entire Earth. More importantly, cities are not predictable (Alexander 1987, Batty 2018), for they should be evolved in a piecemeal fashion, through individual's daily effort to make our environments more beautiful or more living. In this regard, the new approach provides a structural and objective guidance for city development.

The city of Amsterdam presents a good example of how a traditional city was emerged and evolved according to the two underlying laws of living structure. On the other hand, the city of London demonstrates how a traditional city can be decomposed into the substructures of the substructures of substructure and so on, guided by the two design principles. Alexander (2002–2005) demonstrated the 10 simulated cycles of transformations of St. Mark's Square over a 1000-year development period. Every one of these cycles is a wholeness-enhanced transformation, through which the underlying living structure or wholeness is extended, enhanced, and healed. In a similar fashion, the differentiation processes shown in Figures 4 and 7 can also be considered to be wholeness-enhanced transformations. We do not claim that the detected substructures are exactly what happened in the course of evolution, but they are a kind of substructures with respect to growth and form in biology as well as in cities (Thompson 1917). In this connection, the new approach or the theory of living structure in general has taken the understanding of growth and form a step forward towards the making or the design.

We encourage readers of this paper to conduct a thought experiment by extending the living structures studied in this paper. The city of London is a substructure of the Greater London, which is a substructure of the Great Britain, which is a substructure of the European continent, which is a substructure of the Earth's surface, and so on toward the entire universe of the largest scale. The same thought experiment can be carried out in the opposite direction; for example, the leaf vein is the superstructure of the superstructure of the superstructure and so on towards the smallest of the Planck length. In general terms, any living structure is the structure of the structure of the structure and so on. Through the thought experiment, we can achieve a better understanding of the new organismic cosmology (Whitehead 1929), which is shared by many other scientists and philosophers. For example, the eminent quantum physicist David Bohm (1980) conceived and developed a quantum theory of wholeness, which is likened in spirit to Alexander's wholeness.

Back in 1988, Alexander and Bohm spent two days together exchanging their views on the conceptualization of space and the nature of the universe. There are many shared ideas and thoughts among these two great minds. Both believe that the nature of reality is more organismic than mechanistic (Whitehead 1929) governed by the implicate order or something similar. Both believe in the mathematical and physical structure of wholeness (Bohm 1980), in which everything connects everything else, not as a local phenomenon but as a global one. Both believe in the holistic view of reality, not just individual scales but a range of connected scales from the largest to the smallest. Both believe in feelings and consciousness as something real rather than just cognitive things. All these thoughts have deep implications on sustainable urban design. Therefore, through the new developed approach, we can, should, and must adopt the growth way of urban design or planning to make our environments more living or more beautiful towards a sustainable society.

## 6. Conclusion
Built on the growth view of sustainable urban design or planning, this paper develops a new approach to detecting the underlying living structure of an urban environment. Living structure is a physical phenomenon and mathematical structure that is governed by two fundamental laws: scaling law across all scales and Tobler's law on each scale. The new approach takes an urban environment as a coherent structure and decomposes it into many substructures at different levels of hierarchy or scale with far more small scales than large ones. The resulting substructures can be used to quantify or measure the degree of beauty of an urban environment: the more hierarchical levels, the more living or more beautiful it is. Thus, the new approach offers an effective and better measure for understanding and, thereafter, for effectively designing or planning an urban environment to be more living or more beautiful. More importantly, the developed approach can help better understand the kind of problem a city is (Jacobs 1961). A city is the problem of organized complexity, and it has the three fundamental issues: (1) how it looks, (2) how it works, and (3) what it ought to be. Scaling law and Tobler's law both underlie the first two issues, whereas the third issue of what it ought to be can be solved through the two design principles – differentiation and adaptation – for effectively designing or planning an urban environment to be living or more living.

The developed approach provides a holistic perspective on urban environments and is therefore shown to be more powerful than the natural street and axial line models. Previously natural streets or axial lines have been adopted to represent an urban environment as a living structure, for there are far more short natural streets than long ones, or, alternatively, far more short axial lines than long ones. Unlike the detected substructures that constitute a coherent whole, both natural streets and axial lines are still considered to be fragmented entities. For example, natural streets or axial lines cannot be stated recursively. Instead, a living structure can be stated as the structure of the structure of the structure, and so on. Despite their wide applications, both natural street and axial line models – space syntax or topological model – add little insight into sustainable urban design or planning. Many urban design theories are largely for understanding rather than for making or remaking cities. Cities are essentially unpredictable, but they ought to become more living or more beautiful. In this regard, the new approach provides clear guidance through the two design principles: differentiation and adaptation. Our future work points to how the new approach can be applied to urban design or planning in collaboration with design practitioners.

**Highlights:**
- A living structure consists of various substructures at different levels of hierarchy or scale.
- A city is a living structure governed by two fundamental laws and designed by two principles.
- A new approach can help not only understand but also make a city as a living structure.
- The new approach is better than axial line and natural street models.

**Appendix: A Short Tutorial on How to Extract Substructures of a Gray Image Based on ArcGIS**

This tutorial uses a grayscale image as a working example to extract its substructures at different levels of hierarchy or scale. The maple image was freely taken from this website (shorturl.at/hySZ4). Seen from the image (Panel a of Figure A1), the leaf except for the stalk has three levels of hierarchy or scale: the smallest (Panel b) and two larger ones (Panels c and d). The smallest scale has ten substructures shown in blue, while the two larger ones respectively shown in green and red. In the following, we are going to demonstrate in a step-by-step fashion how these substructures can be extracted based on ArcGIS. This tutorial will concentrate on the smallest scale, since the larger ones can be simply aggregated from the smallest substructures.

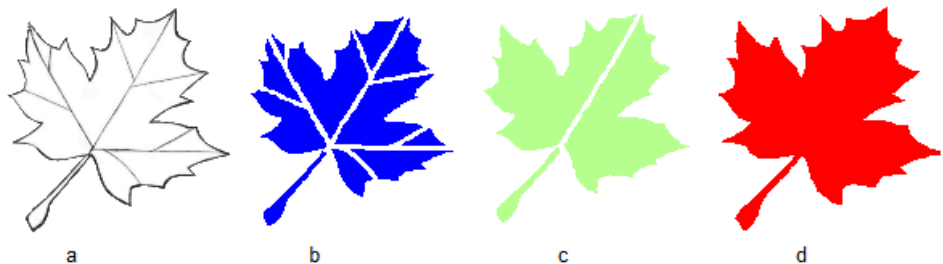

Figure A: The maple leaf image and its substructures at three different levels of hierarchy or scale

The input image is a grayscale image with pixel values usually between 0–255, but this one has values between 10–255 (with values 10 and 255 being the darkest and lightest).

(1) Add the raster image **MapleGray.png** into ArcGIS as a layer and go to ArcToolbox and locate the Reclassify tool (*Spatial Analyst Tools > Reclass > Reclassify*). Reclassify it into two classes using the average value of all the pixels 247 (which is determined by head/tail breaks). Under break values, fill out 247 and 255 (which means pixels in the range 247–255 in one class, while pixels in the range 10–246 is the other class), and save the file as **MapleBinary.tif** and run the reclassify tool. Note that in case the average value 247 cannot well detect the smallest substructures, we can adjust the break value slightly up or down to ensure that the ten smallest substructures are well detected.

(2) Convert the detected smallest substructures from raster to vector using the raster to polygon tool (*Conversion Tools > From Raster > Raster to Polygon*). As input raster use **MapleBinary.tif** and save the file (output polygon features) as **MaplePolygon.shp**. Check the simplify polygons box (if it is not checked) and run the tool.

(3) With smallest substructures detected as above, the larger ones can be obtained by aggregating small substructures into large ones step by step. There are various ArcGIS functions to do that. For example, the aggregate tool (*Cartography Tools > Generalization > Aggregate Polygons*). Input the **MaplePolygon.shp** (with the previously still selected 5 polygons) as Input Features. Save the file as **MapleAggregate.shp** and input a properly determined value (e.g., 15) in the Aggregation Distance field. Uncheck Preserve orthogonal shape if it is not unchecked and run the tool.

Note: If the aggregated output contains holes or has not aggregated everything the distance is most likely too small. You can increase the distance until all selected polygons are aggregated correctly, but make sure that the distance is not too large or too small.